\documentstyle[epsfig,multicol,prb,aps]{revtex}

\begin{document}
\ifx\href\undefined
\else
\errmessage{Don't use hypertex}
\fi

\author{I. I. Mazin$^{a,b}$ and D. J. Singh$^a$}
\address{$^a$Complex Systems Theory Branch, Naval Research Laboratory, Washington,\\
DC 20375-5320 \\
$^b$ George Mason University}
\title{Electronic structure and magnetism in Ru based perovskites}
\date{February 18, 1997}
\maketitle

\begin{abstract}
The magnetic properties of ruthenates with perovskite derived structures,
particularly (Ca,Sr)RuO$_3$ and Sr$_2$YRuO$_6$ are studied within the context
of band structure based Stoner theory. First principles calculations are
used to demonstrate that in all cases the correct magnetic behavior and order
can be obtained without recourse to strong correlation effects, and that the
insulating character of Sr$_2$YRuO$_6$ is reproduced. The different magnetic
states of SrRuO$_3$ and CaRuO$_3$ are shown to be due to the different
structural distortions in these materials, most significantly the larger
rotation of the octahedra in the Ca compound. CaRuO$_3$ is found to be on the
verge of a ferromagnetic instability, leading to the expectation of giant
local moments around magnetic impurities and other anomalous effects in
analogy with fcc Pd metal. Oxygen 2p derived states hybridize strongly with
Ru d states in all three compounds, and O, through this hybridization plays
an unusually large role in the magnetic properties. This involvement of
O is responsible for the strong magneto-structural coupling that is found in
the calculations. Transport properties of CaRuO$_3$ and SrRuO$_3$ are
analyzed using the calculated Fermiology. Unusually large magnon and
paramagnon couplings are found, which are consistent with reported
measurements of the low temperature specific heat  and the resistivity
coefficient.

\end{abstract}

\begin{multicols}{2}
\section{Introduction}

Mixed ruthenates with perovskite based crystal structures have been
receiving considerable attention of late \cite
{1,2,3,4,5,6,7,8,9,10,11,12,13,14,15,16,17,18,19,20,21,22,23,24,25,26,27,27a},
both because of their interesting magnetic properties and because of the
recent discovery of superconductivity in the layered ruthenate, Sr$_2$RuO$_4$
\cite{16}. Despite the rarity of 4d based magnetic materials, SrRuO$_3$ is a
robustly (Curie temperature, $T_C\approx $165 K; magnetization, $m\approx $%
1.6 $\mu _B$/Ru) ferromagnetic metal occurring in a distorted cubic
perovskite structure \cite{28,29,30,31,32}, and $T_C$ can be even further
increased by doping with Pb \cite{13}. However, magnetism
 is easily suppressed by
doping with Ca, although Ca/Sr states are far removed from the Fermi level
and accordingly may not be expected to influence the electronic properties
of SrRuO$_3$ very drastically. Furthermore, Sr$_2$YRuO$_6,$ which has
essentially the same crystal structure as SrRuO$_3,$ but with every second
Ru substituted by Y, is antiferromagnetic, with estimates of the saturation
magnetization even higher than the parent compound ($M\approx 3\mu _B),$
although the critical temperature, $T_N$ is reduced to 26 K. The variety of
magnetic and electronic properties observed in these superficially similar
compounds already poses an interesting theoretical challenge (cf., for
instance, non superconducting cuprates, which despite their large variety,
always show strong antiferromagnetism in Cu-O planes). Besides, there are a
number of interesting observations that deserve attention. These include the
fact that SrRuO$_3$ is the only known ferromagnetic metal among the 4$d$
oxides. As such interesting differences are expected from the much more
abundant 3$d$ oxide magnets. For example, much stronger spin-orbit effects
compared to the 3d systems may be anticipated, and these may manifest
themselves in the magneto-crystalline and magneto-optical properties. In
fact, SrRuO$_3$ does show an abnormally high magneto-crystalline anisotropy
for a pseudo cubic material \cite{32} to the extent that it is difficult to
measure its saturation magnetization using standard measurements of
hysteresis loops, and resulting in some confusion in the older experimental
literature. More recently, Klein and co-workers \cite{5} have measured
strong magneto-optic properties in SrRuO$_3$ epitaxial films. 4$d$ ions
generally have more extended $d$ orbitals than the corresponding 3$d$ ions,
and as a result $4d$ oxides tend to have greater overlap and hybridization
between the transition metal and O 2$p$ orbitals. Besides a tendency towards
greater itinerancy, this can lead to more interplay between structural
degrees of freedom and the magnetic and electronic properties.

As mentioned, additional interest in these ruthenates comes from their
apparent proximity to superconductivity, and possible new insights into the
problem of high-temperature superconductivity that may emerge from their
study. Although the layered perovskite Sr$_2$RuO$_4$ has a modest $T_c$ of
1K (there have been very recent, unconfirmed reports of signatures of
superconductivity at up to 60K in the double perovskite Sr$_2$YRuO$_6$ with
Cu doping \cite{36}), it was suggested that this material may be an
unconventional superconductor. This is based largely on several similarities
with the cuprates: Sr$_2$RuO$_4$ is iso-structural with the first discovered
high-$T_c$ superconductor, shows highly two dimensional electronic
properties, and of course is close to magnetic phases, particularly SrRuO$_3$%
, and Sr$_2$YRuO$_6$. However, the evidence for strong electron correlations
in ruthenates is by far not yet  as compelling as the body that has been
accumulated for the cuprates and many other 3$d$ oxides, and the
question of whether these ruthenates can be treated within the framework of
conventional band theory, or require a strong-correlation based theory,
remains open. 

Several photoelectron spectroscopy experiments have been
reported for Sr$_2$RuO$_4$, which because of its layered crystal structure
is more amenable to such studies than nearly cubic SrRuO$_3$. Yokoya et al. 
\cite{24} and Lu et al. \cite{25}, using angle resolved photoemission (ARPES),
both report observation of Fermi surface sections and extended van Hove
features somewhat like those in the local density electronic structure
calculations, although the positioning of the van Hove singularity relative
to the Fermi energy differs and the dispersion is generally somewhat weaker
than in the calculations, probably due to correlations, but possibly because
of strong electron-phonon and -magnon interactions. Similarly, Schmidt et
al. \cite{21} observed the valence bands of Sr$_2$RuO$_4$ using ARPES, and
found uppermost occupied bands with a width reduced by a factor of 2
compared with band structure calculations. Unfortunately, ARPES is highly
sensitive to the quality of samples and particularly sample surfaces.
Interestingly, polycrystalline but otherwise apparently high quality Sr%
$_2$RuO$_4$ samples can be non-metallic \cite{38}. Angle integrated
photoemission is a more robust technique; using it, Yokoya et al. \cite{23}
found good agreement between the experiment and density functional
calculations, but observe a correlation satellite to the d-band (using
resonant photoemission). Based on these measurements they estimated an
effective Hubbard $U$ of 1.5 eV, which is at least three times smaller than
similar estimates for the cuprate superconductors, casting some doubt on
suggestion that Sr$_2$RuO$_4$ and related ruthenates are very strongly
correlated.

One of the most decisive arguments in favor of the importance of strong
correlations in high-$T_c$ cuprates is the failure of conventional
local-density-approximation band structure calculations to describe even
qualitatively the antiferromagnetism in the undoped parent compounds.
Similarly, the key question for these ruthenates may be which approximation,
strongly correlated or band structure based, is best suited to explaining
the variety of magnetic properties. One of the main purposes of this work is
to determine whether a (similar to the cuprates) failure of the
conventional, mean-field type, band calculations, is present in these
ruthenates.

Within a strong-correlation scenario, the ferromagnetism in metallic SrRuO$%
_3,$ results from the double exchange mechanism, while antiferromagnetism in
insulating Sr$_2$YRuO$_6$ is due to superexchange via two (unlike the 3d
oxides and Cu perovskites) oxygen ions. This is appealing because in the
Mott-Hubbard picture the main factor controlling the magnetic properties is
carrier concentration, which is indeed different in those two materials: in
(Sr,Ca)RuO$_3$ ruthenium is four-valent, that is, its $d$-band is populated
by 4 electrons, while in Sr$_2$YRuO$_6$ the nominal valency of Ru is 5, and
the number of $d$ electrons is 3. On the other hand, integer occupancy does
not favor the double exchange scenario, and, besides, it is unclear how the
Mott-Hubbard model provides a mechanism for suppressing magnetism in CaRuO$%
_3.$ Finally, as we discuss in detail below, conventional band theory in all
the cases we test does yield the correct magnetic ground state, in contrast to
the cuprates and similar correlated 3$d-$oxides. Thus, contrary to some
recently suggested superconductivity scenarios based on strong correlations%
\cite{26,37}, it seems likely that if strong correlations play some role, it
is more of a quantitative than of a qualitative nature.

On the other hand, we note that even if Sr$_2$RuO$_4$ and other ruthenates
are not strongly correlated, the superconductivity could still be
unconventional, for instance arising from a magnetic mechanism. In this
regard, a number of measurements indicate anomalously large scattering of
electrons by spin fluctuations\cite{15}, complicated by a strong
magnetoelastic coupling\cite{2}. Cyclotron masses\cite{22}, the specific
heat, and the paramagnetic susceptibility \cite{16} are all strongly
renormalized. While there is always the possibility of ascribing this
renormalization to  strong correlations, the simplest explanation  may be
strong electron-phonon-magnon interactions. Further, an abnormally large
transport coupling constant, $\lambda _{tr}$ is required to rationalize the
temperature dependence of the resistivity with the calculated Drude plasma
energies\cite{18}, although this value is consistent with the specific heat
enhancement. Unusual temperature dependencies of the Hall effect \cite{20} 
were found in CaRuO$_3$ and in SrRuO$_3$. We shall return to the transport
properties later in the paper; it is plausible that they can be reconciled
with the conventional one-electron mechanism, despite the unusual $T-$%
dependences.

The main purpose of the present paper is to study the magnetic phases and
the relative importance of correlation and band structure effects for
obtaining the magnetic properties. We focus on the double perovskite Sr$_2$%
YRuO$_6$ and the ferromagnetic-paramagnetic transition in (Sr,Ca)RuO$_3$
with increasing Ca content, and we shall show that conventional band theory
is fully able to describe the variegated magnetic properties in this
family of materials.

\section{First Principles Calculations}

\subsection{Structure, Magnetism and Ionic Considerations:}

As mentioned, SrRuO$_3$ occurs in  an orthorhombic, Pbnm, GdFeO$_3$
structure, which has four formula units per cell. It is interesting to note
that this is the same generic structure as LaMnO$_3$ and related manganites
that have received considerable recent attention because of the discovery of
colossal magnetoresistance effects in some of these. Further SrRuO$_3$ has
the same nominal $d$ electron count as LaMnO$_3$, although unlike LaMnO$_3$
it is a ferromagnetic metal even without doping. In LaMnO$_3$ the distortion
from the ideal cubic perovskite crystal structure
consists of both rotations of the O
octahedra and Jahn-Teller distortions of them to yield Mn-O bond length
variations of more than 10\%. This is understood in ionic terms as a
result of the fact that the high spin Mn ion with this electron count has a
half full majority spin e$_g$ orbital favoring a Jahn-Teller distortion. In
contrast, SrRuO$_3$ occurs with a reduced magnetic moment and its distortion
consists of almost rigid rotations of the O octahedra with practically no
accompanying variations in the Ru-O bond lengths.

CaRuO$_3$ occurs in the same crystal structure and symmetry as  SrRuO$_3$,
also with no
evident Jahn-Teller distortion of the O octahedra, but with approximately
twice larger rotations. Such rotations 
are common in perovskite based materials and are usually understandable in
terms of ionic size mismatches between the A and B site cations. Such an
explanation is consistent with the trend observed in (Sr,Ca)RuO$_3$ since
the Ca$^{2+}$ ionic radius is approximately 0.15 \AA\ smaller than Sr$^{2+}$%
. Although CaRuO$_3$ is paramagnetic, it is believed to be rather close to
magnetic instability.

Sr$_2$YRuO$_6$ is an antiferromagnetic insulator that occurs in a distorted
but well ordered double perovskite structure. This is derived from the
perovskite, SrRuO$_3$ by replacing every second Ru by Y, such that the
remaining Ru ions form an fcc lattice. The structural units are thus Ru-O
and Y-O octahedra, with the Sr ions in the A site positions providing charge
balance. Each Ru-O octahedra shares a single O atom with each neighboring
Y-O octahedra, and vice versa, but there are no common O ions shared between
 different
Ru-O octahedra. The primary distortions from the ideal perovskite derived
structure, consist of (1) a substantial breathing of the octahedra to
increase the Y-O distance to 2.2 \AA {} at the expense of the Ru-O distances
which become 1.95 \AA {} and (2) rotations of the octahedra to reduce the
closest Sr-O distances, consistent with the ionic sizes.
These distortions reduce the symmetry to monoclinic
(P21/n). A related view of the crystal structure is based on the fact that
Y, like Sr is fully ionized in such oxides, and accordingly is a spectator
ion providing space filling and charge to the active Ru-O system but playing
no direct role in the electronic or magnetic properties. From this point of
view, Sr$_2$YRuO$_6$ consists of independent rigid, but tilted, (RuO$_6$)$%
^{7-}$octahedral clusters, arranged on a slightly distorted fcc lattice.
Hopping then proceeds between Ru ions in neighboring RuO$_6$ clusters via
two intervening O ions

Since Y is tri-valent, the Ru is formally 5-valent ($4d^3$) in this compound
instead of formally tetra-valent as in perovskite SrRuO$_3$. In the
octahedral crystal field, the Ru $t_{2g}$ orbitals lie below the $e_g$
orbitals, so that in the high spin state the majority spin Ru $t_{2g}$
manifold would be fully occupied, and all other Ru 4d orbitals unoccupied.
This Jahn-Teller stable configuration is consistent with the experimental
observation that the bond angles and bond lengths within the Ru-O octahedra
are almost perfectly equal, but the Ru moment of 1.85 $\mu _B$/Ru measured using
neutron diffraction is considerably smaller than the 3  $\mu _B$/Ru that
would be expected in the high-spin configuration.

First principles studies of SrRuO$_3$ have shown that its electronic structure
involves rather strong Ru-O covalency, and that O $p$-derived states
participate substantially in the magnetism and the electronic structure near
the Fermi energy, which is important for understanding the transport
properties. As will be discussed below, a similar covalency is present in
CaRuO$_3$ and the differences in the magnetic ground states of CaRuO$_3$ and
SrRuO$_3$ are due to band structure effects related to the modulation of the
Ru-O hybridization by the structural distortion. In this regard, it should
be noted that Ru$^{5+}-$O hybridization may be even stronger in Sr$_2$YRuO$_6
$, based on the expectation that the O $2p$ manifold would be even higher in
energy with respect to the Ru $d$ states. The similar Ru-O distances in SrRuO%
$_3$ and Sr$_2$YRuO$_6$ (less than 0.03 \AA {} longer than in Sr$_2$YRuO$_6$%
) and the fact that 5 is not a common oxidation state for Ru also suggest
strong covalency in the double perovskite. Here we report density functional
calculations of the electronic and magnetic properties of Sr$_2$YRuO$_6$.
These confirm the strongly hybridized view of these materials and provide an
explanation for the electronic and magnetic properties.

\subsection{SrRuO$_3$}

The electronic structure of SrRuO$_3$ has been described elsewhere\cite{6,27}.
Here we repeat, for completeness, the main results, and also discuss some
quantitative differences between the published calculations.

There have been two recent band structure calculations for SrRuO$_3$\cite
{6,27}. In both works the calculations were performed for both an idealized
cubic perovskite structure and the experimental crystal structure. Allen 
{\it et al}\cite{27} interpreted their experimental measurements in terms of
 the
band structure calculated within the  local spin density approximation
 (LSDA) using the linear muffin-tin orbitals (LMTO) method. Singh \cite
{6} used the general potential linearized augmented plane-waves (LAPW)
method to calculate  electronic and
magnetic properties. The two studies yielded
reasonably similar results for the electronic structures near the Fermi
energy although some noticeable differences are present. Important for
interpreting 
experimental results
 are the differences in the density of states and in the Fermi
velocities. The latter were found in Ref. \onlinecite{6} to be almost
isotropic, while in Ref.\onlinecite{27} strong anisotropy of the Fermi velocity
(about 30\% in each channel) was reported. The ratio $N_{\uparrow
}/N_{\downarrow }$
found in Ref. \onlinecite{27} is 50\% larger than that in Ref. 
\onlinecite{6}. Most
important, the overall shape of the density of states within a $%
\pm 0.2$ Ry window at the Fermi level is rather different. It is known that
the accuracy of the 
 atomic sphere approximation calculations can be difficult to control  for
materials with open crystal structures and low site symmetries due to
sensitivity to the computational parameters (e.g., basis set,
inclusion of empty spheres in lattice voids, linearization parameters etc.).
Since we wanted to use LMTO-ASA technique in analyzing the calculated band
structure, we have repeated the LAPW calculations reported in Ref.
\onlinecite{6}
using a standard LMTO-ASA package {\it Stuttgart-4.7}. We found it necessary
to include 10 empty spheres per formula unit to achieve adequate space
filling in the distorted structure (in the cubic perovskite structure this was
not needed). The result appeared to be much closer to the LAPW results of
Ref. \onlinecite{6} than to the LMTO
ones of Ref. \onlinecite{27}; Ref. \onlinecite{27} does
not mention use of any empty spheres, in which case insufficient space
filling could have influenced the calculation.
The results given here are from LAPW calculations,  except
where specifically noted otherwise.

Calculations for SrRuO$_3$ in the ideal perovskite structure yielded a spin
moment of 1.17 $\mu _B$ per formula unit, while calculations including the
experimentally observed rotations yielded a larger moment of 1.59 $\mu _B$
in accord with recent experimental results. Only a portion of the total
moment resides on the Ru sites (64\% in the LAPW MT sphere, and 67\% in
the LMTO atomic sphere).
The electronic density of states has a gap in the spin majority channel
which is only 20 mRy above the Fermi level.  The fact 
that SrRuO$_3$ is so
close to a half-metal is important for understanding its transport
properties, and  the fact
 that they are so sensitive to magnetic ordering
(and, correspondingly, to temperature).

\subsection{CaRuO$_3$}

\begin{figure}[tbp]
\centerline{\epsfig{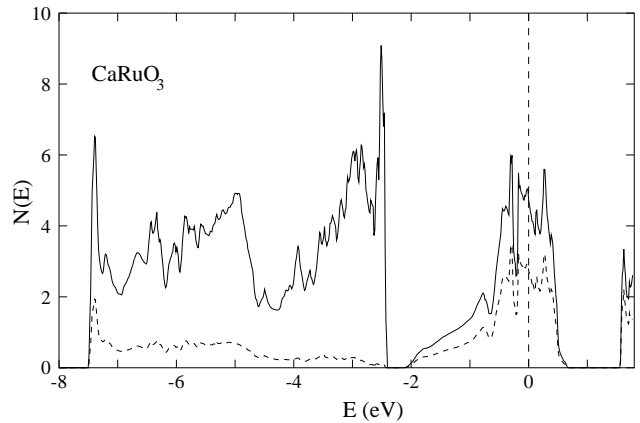}}
\vspace{0.1in} \setlength{\columnwidth}{3.2in} \nopagebreak
\caption{LAPW density of states of CaRuO$_3$ in its actual crystal
structure. The  total density of states is shown by the solid line, Only Ru(d)
partial density of states is  shown (dashed line), because the O(p) density is
approximately the difference between the total and the Ru(d) densities.
Here and in the other figures all densities  of states are per spin and
formula unit. }
\label{CR3-DOS} \end{figure}

Experimentally, CaRuO$_3$ is a paramagnetic metal. This fact suggests that
the rotation of RuO$_6$
the octahedra is antagonistic to magnetism (since larger rotations constitute
the main structural difference between CaRuO$_3$ and SrRuO$_3$). However, this
conjecture is apparently at odds with the calculated result that the
equilibrium magnetization in SrRuO$_3$ is smaller in an ideal cubic
perovskite 
structure than in the actual
distorted one. As a first step to understanding this,
we have extended our calculations to CaRuO$_3$ in its experimental
structure. Details of the method are as in Ref.\onlinecite{6}. 
The resulting density of states is shown in Fig.\ref{CR3-DOS}.
We find that indeed the magnetism is suppressed in this case,
though in a very borderline fashion. Fixed spin moment calculations of the
total energy as a function of spin magnetization for CaRuO$_3$ show a very
extended flat region, extending to near 1.5 $\mu _B$ per formula unit. This
is reminiscent of fcc Pd which also shows such a flat region. This
borderline state implies a high spin susceptibility and explains the fact
that low doping can induce a ferromagnetic state. Further, para-magnon like
spin excitations should be very soft in this material and magnetic
impurities may be expected to induce giant induced local moments. There are
already some reports that this is the case in CaRuO$_3$ \cite{crow}.

Having shown that the ferromagnetism in SrRuO$_3$ and its suppression in
CaRuO$_3$ can be described using band structure methods, we turn to the
question of why these two perovskites have different magnetic properties. To
determine whether the key difference between the materials is structural we
have performed calculations for CaRuO$_3$ using the crystal structure of
SrRuO$_3$. These calculations yield a spin magnetization of 1.68 $\mu _B$
per formula unit and a magnetic energy of 0.06 eV/Ru, very similar to SrRuO$%
_3$.
Calculations for the intermediate structure formed by a linear average of
the experimental CaRuO$_3$ and SrRuO$_3$ structures, yield a similar spin
moment of 1.53 $\mu _B$ per formula unit and a magnetic energy of only 0.029
eV/Ru (note the similarity of the magnetizations and large variation of
magnetic energy). To within the accuracy of our calculations this
paramagnetic - ferromagnetic energy difference becomes zero just at the
experimental CaRuO$_3$ structure. Since ferromagnetism in the (Sr,Ca)RuO$_3$
is apparently strongly coupled to the rotation of the octahedra, alloying
the A site cation is expected to be an effective means for tuning the
magnetic properties. Alloying CaRuO$_3$ with larger divalent cations should
generally induce ferromagnetism while alloying SrRuO$_3$ with smaller
cations should suppress ferromagnetism. BaRuO$_3$, while a known compound,
occurs in a different crystal structure and is not magnetic. However, Pb can
be partially substituted on the Sr site, and it is known that introduction
of this slightly larger divalent cation does increase $T_C$ in SrRuO$_3$.

Later in the paper we shall analyze the transformation of the band structure
of (Sr,Ca)RuO$_3$ upon increase of the tilting in more detail and will show
that the nonmonotonic dependence of the equilibrium magnetization on
tilting is a straightforward consequence of a natural evolution of the band
structure near $E_F$ with the structural distortion.

\subsection{Sr$_2$YRuO$_6$}

The electronic and magnetic structure of Sr$_2$YRuO$_6$ was calculated using
the full experimental crystal structure of Battle and Macklin \cite{SYR-str}
except that the very small (0.23\%) lattice strain was neglected. Additional
calculations were performed for idealized structures neglecting the tilting
of the octahedra to help understand the role of this distortion, which
changes the angles and distances along the Ru-O-O-Ru hopping paths. These
local density approximation calculations were performed using the general
potential LAPW method \cite{LAPW} including
local orbital extensions
 \cite{LAPW1} to accurately treat the O $2s$ states
and upper core states of Sr and Y as well as to relax any residual
linearization errors associated with the Ru $d$ states. A well converged
basis consisting of approximately 2700 LAPW basis functions in addition to
the local orbitals was used with O sphere radii of 1.58 a.u. and cation
radii of 2.10 a.u. This self-consistent approach has a flexible
representation of the wavefunctions in both the interstitial and sphere
regions and makes no shape approximations to either the potential or charge
density. As such it is well suited to materials with open structures and low
site symmetries like Sr$_2$YRuO$_6$. In addition, we used 
the LMTO method  in the atomic sphere approximation and
tight-binding representation\cite{LMTO} (Stuttgart code, version 4.7)
to get better insight in the calculated electronic structure.
 The LMTO-ASA method is less
accurate than the full-potential LAPW but it provides more flexibility in
the way how the results are represented and how they can be analyzed in 
tight-binding language.

\begin{figure}[tbp]
\centerline{\epsfig{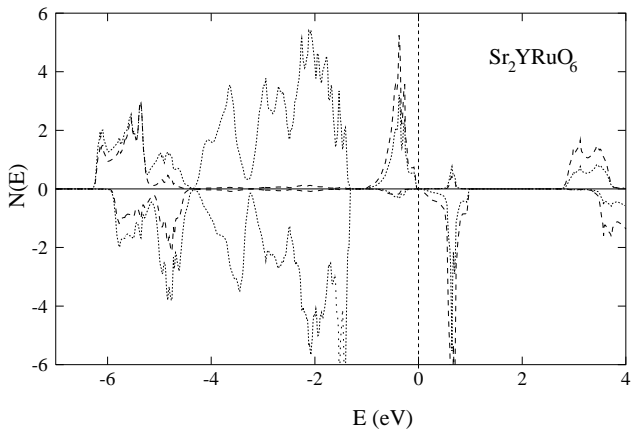}}
\vspace{0.1in} \setlength{\columnwidth}{3.2in} \nopagebreak
\caption{LAPW density of states of antiferromagnetic Sr$_2$YRuO$_6$.
Partial densities of states of Ru(d) and O(p) orbitals are shown
by the dashed and dotted lines, respectively.} 
\label{DOS-SYR-AF} \end{figure}

Calculations were performed at the experimental structure for ferromagnetic
(F) and the observed antiferromagnetic (AF) orderings. The AF ordering is
0.095 eV/Ru lower in energy than the F ordering, and has an insulating gap
in the band structure, consistent with the experimental ground state. The
insulating gap of 0.08 eV is between majority and minority spin states and
may yield only a weak optical signature. The Ru moment
as measured by the magnetization within the Ru LAPW sphere is 1.70 $\mu _B$
for the AF state and 1.80 $\mu _B$ with the F ordering, in reasonable accord
with the neutron scattering results The similar moments with different spin
configurations  suggests that a local moment picture
of the magnetism is appropriate for Sr$_2$YRuO$_6$. This is in contrast to 
perovskite
SrRuO$_3$. Similar to SrRuO$_3$
there are substantial moments within the O LAPW spheres as well as the Ru
spheres, amounting to approximately 0.10 $\mu _B$/O (AF ordered) and 0.12 $%
\mu _B$/O (F ordered). These cannot be understood as tails of Ru 4d orbitals
extending beyond the LAPW sphere radii, since such an explanation is
inconsistent with the radial dependence of these orbitals, but rather they
arise
from polarization of the O ions due to hybridization, which is evidently
strong both from this point of view and from the calculated electronic
structure, discussed below. The total local moment per formula unit is of
mixed Ru and O character and amounts to 3 $\mu _B$/f.u., which is
approximately 60\% Ru derived and 40\% O derived (the interstitial
polarization of 0.5 - 0.7 $\mu _B$/cluster derives from both Ru and O, but
is assigned as mostly O in character based on the extended 2p orbitals of
negative O ions and the small O sphere radius, and results of LMTO-ASA
calculations which do not have any interstitial volume). The calculated
exchange splittings of the O 1s core levels are 80 to 95 meV depending on
the particular O site. The O polarizations may be observable in
neutron experiments if O form factors are included with Ru in the
refinement. Such an experiment is strongly suggested by the present results.

\begin{figure}[tbp]
\centerline{\epsfig{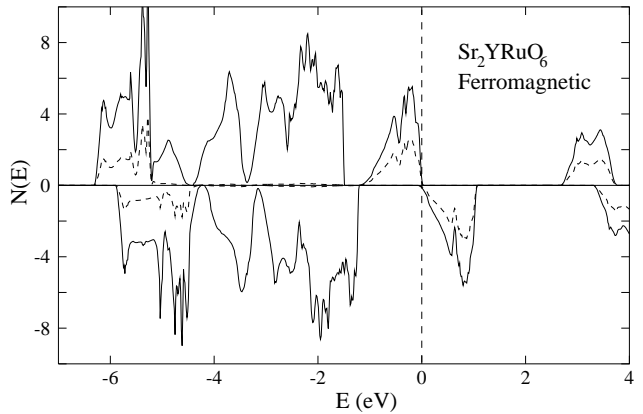}}
\vspace{0.1in} \setlength{\columnwidth}{3.2in} \nopagebreak
\caption{LAPW density of states of ferromagnetic Sr$_2$YRuO$_6$.
Partial density of states of Ru(d) orbitals and the total density
of states are shown
by the dashed and solid lines, respectively.} 
\label{DOS-SYR-F}
\end{figure}

Projections of the electronic density of states (DOS)  of antiferromagnetic
Sr$_2$YRuO$_6$  onto the LAPW spheres
are shown in Fig. \ref{DOS-SYR-AF}, where majority and minority spin
projections onto a Ru ion and the six O ions in its cluster are shown. The
DOS in two spin channels are similar in shape apart from an exchange
splitting throughout the valence energy region and shows evidence of a strongly
hybridized electronic structure. The details of this structure are deferred
to the tight binding analysis below, except to mention the exchange
splitting of the essentially pure O 2p states between -4 and -6 eV relative
to the Fermi energy ($E_F$) and the fact that there are substantial Ru 4d
contributions to the minority spin channel between -4 and -6 eV as well as O
contributions above $E_F$ implying that the average Ru 4d occupancy is
considerably higher than d$^3$. Although assigning charge in a crystal to
various atoms is an ambiguous procedure, integration of the d-like DOS
implies an average near d$^5$ similar to perovskite SrRuO$_3$. The magnetic
moments derive from polarization of three bands near $E_F$ by an exchange
splitting of 1 eV. The F ordered DOS (Fig. \ref{DOS-SYR-F}) is very similar
to that in the AF state, but the exchange splitting is somewhat smaller and
the bandwidth somewhat larger, resulting in a slight semimetallic overlap of
majority and minority spin bands at $E_F$ which reduces the spin moment from
3.0 to 2.97 $\mu _B$/f.u..

Parallel calculations were performed using a
structure in which the tilting of the RuO$_6$ clusters is suppressed. As
with the actual experimental structure, the AF ordering is lower in energy
than the F ordering. However, in this case the band structures are metallic
for both orderings, showing that the tilting is crucial for the insulating
state. As will be discussed below there is a substantial coupling
between the magnetic order and this structural degree of freedom.

\section{Tight-binding interpretation and physical properties}

\subsection{Sr$_2$YRuO$_6$}

\subsubsection{Single RuO$_6$ cluster\label{1cluster}}

Somewhat unexpectedly, the easiest compound to understand is the Sr$_2$YRuO$_6$
double perovskite. Sr and Y, as is common in perovskites, are fully ionic,
so that the states around the Fermi level barely have any Sr or Y character.
Thus, as mentioned, this compound can be viewed as consisting of rigid RuO$%
_6 $ octahedra, arranged on an fcc lattice, and loosely connected to each
other. We will show below that this intuitive picture provides very good
qualitative and quantitative interpretation of the full-scale band structure
calculation. In contrast with Sr$_2$RuO$_4$ or Sr$_x$Ca$_{1-x}$RuO$_3,$ no
octahedra share oxygens. The octahedra are slightly tilted, which we shall
neglect for the moment (the effect of tilting is in a certain sense important
and will be discussed later). Accordingly, we begin by discussing a single
cluster.

\begin{figure}[tbp]
\centerline{\epsfig{file=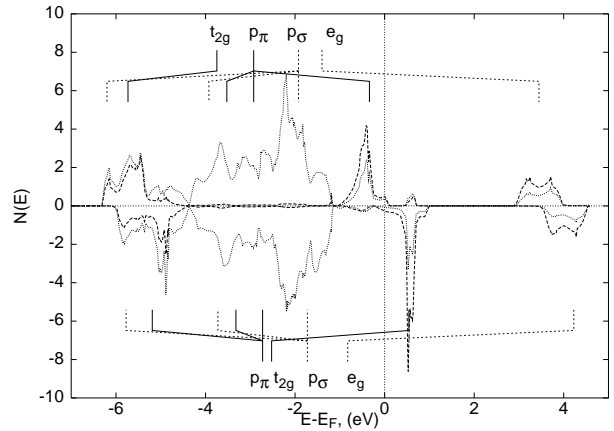,width=0.95\linewidth}}
\vspace{0.1in} \setlength{\columnwidth}{3.2in} \nopagebreak
\caption{Calculated LAPW density of states
 for cubic Sr$_2$YRuO$_6$ (antiferromagnetic)
 and the level scheme for an individual RuO$_6$ cluster. Notations for 
the density of states are the same as in Fig. \protect\ref{DOS-SYR-AF}.
$T_{2g}$ levels and their parent states are shown by solid lines, the
$E_{g}$ levels and states by dashed lines.  Compared with the formulas in
Section \ref{1cluster}, additional small O-O hoppings $\tau _\pi $
and $\tau _\sigma $ are taken into account; These split off the non-bonding
levels mixed oxygen states with $E_0(p_\sigma )-4\tau _\sigma $ and $%
E_0(p_\pi )-2\tau _\pi $.} \label{levels} \end{figure}

The electronic structure of a single RuO$_6$ cluster is governed by the
relative position of Ru $d$ and O $p$ levels, and the corresponding hopping
amplitudes. The Ru $d$ states are split by the crystal field into two
manifolds consisting of 3 $t_{2g}$ and 2 $e_g$ levels, respectively, and
these are separated by $\approx 1$ eV. The O $p$ levels are subject to a
crystal field splitting at least three times smaller, and yield 9 $p_\pi $
states, which form $pd\pi $ bonds with Ru, plus three $p_\sigma $ states,
which participate in the $pd\sigma $ bonding. After including $pd$ hopping,
the system of levels becomes, for each spin channel: 13 nonbonding: 4$\times
E_0(p_\sigma )+9\times E_0(p_\pi ),$ 5 bonding: 2$\times E_{-}(E_g)+3\times
E_{-}(T_{2g}),$ and 5 antibonding: 2$\times E_{+}(E_g)+3\times E_{+}(T_{2g}),
$ where $E_0$ are pure ionic levels, and $E_{\pm }(E_g)=0.5\{E_0(p_\sigma
)+E_0(e_g)\pm \sqrt{[E_0(p_\sigma )-E_0(e_g)]^2+16t_\sigma ^2}\},$ $E_{\pm
}(T_{2g})=0.5\{E_0(p_\pi )+E_0(t_{2g})\pm \sqrt{[E_0(p_\pi
)-E_0(t_{2g})]^2+16t_\pi ^2}.$ The actual ordering of levels in RuO$_6,$ as
shown on Fig.\ref{levels}
is $E_{-}(T_{2g})\approx E_{-}(E_g)<E_0(p_\sigma
)<E_0(p_\pi )<E_{+}(T_{2g})<<E_{+}(E_g)$. The last inequality leads to a
substantial gap ($>$ 2 eV) between the antibonding $T_{2g}$ and the 
antibonding $%
E_g$ bands in the solid. This large gap is only partially due to the crystal
field, and arises largely from the stronger (relative to $pd\pi )$ $pd\sigma 
$ bonding. The exchange splitting is, naturally, weaker than this enhanced
crystal field splitting, and the Hund's rule does not apply to the
high-lying antibonding $E_g$ states, which remain empty in both spin
channels. Neglecting those states, there are 21 levels to be occupied by 39
valence electrons. Here Hund's rule does apply and tells us to populate all
21 spin-majority states, and all but the three antibonding $T_{2g}$ levels
in the spin-minority channel. Thus for the electronic properties of
the crystal these six spin-up and spin-down  $T_{2g}$ 
 molecular orbitals are of primary relevance; the
symmetry of these orbitals is the same as for $d(t_{2g})$ states in a
transition metal ion. We now use this information to analyze the electronic
structure of crystalline Sr$_2$YRuO$_6$.

\subsubsection{Intercluster hopping and exchange.}

When a solid is built out of the clusters, the molecular levels broaden into
bands, which however remain quite narrow in this material. Although the main
intercluster hopping occurs via $dd\sigma $ matrix elements (here and below
we mean for $d$ the Ru-O molecular orbitals with the effective $d-$%
symmetry), the intercluster distance is large and the effective hopping
amplitude is
small. Thus one may conjecture that the valence band formed out of the majority
spin molecular $T_{2g}$ orbitals, and the corresponding minority spin band
do not overlap, and that the crystal, in either the ferro- or
antiferromagnetic states, remains insulating. A more detailed analysis,
as discussed later in the paper, reveals a difference between the
ferro- and antiferromagnetic ordering, namely that the bandwidth
is slightly larger, and the exchange splitting slightly smaller in the
former case. In fact, our LDA calculations,
described above, yield an insulating antiferromagnetic ground state, with a
small gap of about 0.07 eV; they also give
 a metastable semimetallic ferromagnetic
state, with a band overlap of a few meV. 

 Let us now analyze this band
structure in the tight-binding terms.
A nearest neighbor model should be a good starting approximation. Let us
begin with the ferromagnetic case, and consider the undistorted crystal
structure (no tilting of oxygen octahedra). The main parameter is now the $%
xy-xy$ hopping amplitude, $\tau _\sigma =0.75t_{dd\sigma }.$ In the nearest
neighbor approximation, the three $T_{2g}$ bands do not hybridize with each
other. Each of them, however, disperses according to $E_k=$ $%
E_{+}(T_{2g})+4\tau _\sigma \cos (k_xa/2)\cos (k_ya/2),$ and the
corresponding permutations of $x,y,z.$ Including  $dd\pi $ hopping, the
bands hybridize among themselves, resulting in a
 further increase in the bandwidth. The
calculated LDA bands have widths of approximately
 1.1 eV, corresponding to $\tau
_\sigma \approx 0.14$ eV. $dd\pi $ hopping effects are responsible for the
deviations from the dispersion above by about 0.1 eV. Importantly, there is
no repulsion between the valence bands and the conduction bands, because
they are fully spin polarized with the opposite spins. This situation
changes, however, in the antiferromagnetic case.

The observed magnetic ordering corresponds to ferromagnetic 001 planes
stacked antiferromagnetically. Each RuO$_6$ cluster has thus 4 neighbors
with the same and 8 neighbors with the opposite spin. Correspondingly, of
three $T_{2g}$ derived bands one ($xy$) remains essentially the same as in
the ferromagnet, and two other loose their dispersion to the first order in $
\tau _\sigma ,$ since the relevant neighboring clusters have only states of
the opposite spin at this energy. Instead, for those bands there is a
hybridization between the valence and the conduction bands, because now the
orbitals with the same spin on the neighboring clusters belong to these
different bands. The hybridization matrix element is $t_\sigma ({\bf k)=}%
4\tau _\sigma \cos (k_{x,y}a/2)\cos (k_za/2),$ and produces an additional
bonding energy $2J\approx \langle t_\sigma ^2({\bf k)\rangle }/\Delta $ per
cluster ($\Delta \approx 1$ eV is the exchange splitting). This yields about
0.08 eV which is very close to the calculated LDA energy difference (0.12
eV) between the AFM and FM configurations in the ideal 
undistorted structure. In
the actual crystal structure the
oxygen octahedra are tilted by about 12$^{\circ }$, so
that $\tau $ is reduced by about 15\% (neglecting $\tau _\pi $ etc.),
yielding $2J\approx 0.06$ eV. Our first-principles LAPW calculations give
for the bandwidth approximately 0.9 eV, that is $\tau _\sigma \approx 0.11$
eV, and $2J\approx 0.05$ eV. The calculated LAPW energy difference is 0.095
eV, the same reduction from the undistorted case as given by the simple
tight binding estimate above. It is worth noting that while
this mechanism gives an effective antiferromagnetic exchange interaction $%
J\propto \tau ^2/\Delta ,$ the underlying physics is very similar to, but not
identical with, the usual superexchange interaction in 3$d$ oxides, $J\propto
t^2/U $. The differences are that instead of metal-oxygen-metal hopping here
the relevant hopping is direct cluster-cluster hopping and the energy
denominator is the band gap due mainly to intracluster
exchange, rather than Coulomb correlations described by a Hubbard $U$.

One can also estimate the Neel temperature, using the above value for $J.$
To do that, let us begin with noting that this system represents a very good
approximation to the antiferromagnetic nearest neighbor fcc model. The
strong magnetoelastic coupling discussed below does not favor non-collinear
 spin configurations,
so the direction of the cluster magnetic moments is fixed. The
magnetic coupling $J_2$ with the next nearest neighbors can be safely
neglected. Indeed, it is governed by the $dd\sigma $ hopping. Although $\tau
_\sigma $ is larger than $\tau _\pi ,$ usually by a factor of the order of
2, the larger distance, in the canonical scaling $d^{l+l^{\prime }+1}$ gives
a factor of $2^{-2.5}=0.18,$ and the energy denominator in the equation for $%
J$ is about 10 times larger. Taken together, one expects $J_2$ to be at
least two order of magnitudes smaller than $J.$ The antiferromagnetic fcc
Ising model is well studied\cite{leib}. Despite magnetic frustration, it has
a Neel temperature of approximately
 1.76$J,$ for the spin 1/2 and  approximately 1.33 $J,$ for the spin 1,
which in our case corresponds to
700--900 K. The measured $T_N$ is 26 K, in apparent severe disagreement
 with our estimate.

It is tempting to ascribe this to intracluster Hubbard-like correlation
effects, which can increase
the gap and reduce $J.$ Moreover, since the $t_{2g}$ band width is only 1
eV, even a moderate Hubbard repulsion could  affect $J$. One can get a very
rough upper estimate this effect as follows: The energy of the Coulomb
repulsion of two electrons placed in two $t_{2g}$ orbital on the same
cluster is (assuming about equal population on Ru and O) $U\approx
0.25U_{O-O}+0.25U_{Ru-Ru}+0.5U_{Ru-O},$ where $U_{O-O}$ is the Coulomb
repulsion of two electrons localized on two neighboring oxygens etc.. $%
U_{O-O}\approx 1/d_{O-O}=4.4$ eV; $U_{Ru-Ru}$ is believed to be about 1.5 eV%
\cite{24}, and for $U_{Ru-O}$ we use 3 eV, keeping in mind that the
charge-transfer metal-oxygen energy for the 3d oxides is about 4.5 eV and
the metal-oxygen distance is 50\% smaller there. Then, we arrive at $U<3$
eV. It is unclear to what extent this $U$ will be reduced by screening by
surrounding cluster and by intracluster charge redistribution, but this
effect would definitely be substantial. Anyway, using 3 eV as a very safe
upper bound, we get for the lower bound on 2$J$ approximately 0.03 eV, which
corresponds to $T_c$ of at least 300 K. Thus, strong correlations alone cannot
explain anomalously low Neel temperature of this compound. Another
possibility 
to reduce the transition temperature is magnetoelastic coupling,
which is subject of the next section.

\subsubsection{Magnon-phonon coupling}

The fact that magnetic excitations and phonons are coupled in ruthenates is
known\cite{2}, but not well understood from a microscopic point of view.
In the case of Sr$_2$YRuO$_6$ it is, however, reasonably clear: with
increasing tilting angle the $\tau _\sigma $ hopping must decrease and
with it the antiferromagnetic stabilization energy and effective exchange
constant $J.$ This is confirmed by our first-principle results.
In other words, magnetic excitations flipping
the spin of a RuO$_6$ cluster are coupled with this phonon mode, changing
the tilting angle (which is the soft mode for the transition from the cubic
structure into the tilted one). A dimensionless coupling constant may be
defined as $\lambda =d\ln J/dQ,$ where $Q=u_O\sqrt{2M_O\omega /\hbar }$ is
the phonon coordinate. Here $u_O$ is the displacement of oxygens from their
equilibrium positions,
and $\omega $ is the frequency of the phonon. Very roughly, $%
M_O\omega ^2=8\Delta E/d^2,$ where $\Delta E$ is the energy difference
between the cubic and the distorted structure, taken per one oxygen, and $d$
is the equilibrium oxygen displacement. From our calculations, $\Delta E=90
$ meV. Experimentally, $d\approx 0.4$ \AA . Thus, $\omega \approx 270$ cm$%
^{-1}$. Now, using $2J\propto \tau _\sigma ^2\propto \cos ^22\theta ,$ where 
$\theta $ is the tilting angle, we can estimate $d\ln J/du_O\approx 8\theta
_0^2/d\approx 0.8$ \AA $^{-1}.$ In fact, linear interpolation of $J$
between the
cubic and equilibrium structure gives the same number for $d\ln J/du_O.$
Thus $\lambda $ is about 0.17 for this phonon mode, which means that the
characteristic (e.g., zero-point motion) amplitude of the librations of the
octahedra around their equilibrium position will produce sizable changes in
the effective exchange constant. The thermodynamics of such a system is
interesting and unusual, but its discussion goes beyond the scope of this
paper. It is important to note, however, that the long-range 
order in the nearest-neighbors antiferromagnetic
FCC Ising model appears exclusively because
of the finite temperature entropy contribution to the free energy\cite{mac}.
While at $T=0$ there is an infinite number of degenerate states, ordered in
two dimensions and disordered in the third one, at $T>0$ this degeneracy
is lifted because of different spectra of low-energy spin-flip excitations
in the different ground states. As long as such spin-flip excitations are 
coupled with the phonons, the standard consideration of the AFM FCC Ising
model does not apply, and the transition is not necessarily at $T\agt J$.
However, the long-range two-dimensional AFM correlations should be present
up to $T\approx J$, and could in principle be seen in some experiments.

\subsubsection{Extended Stoner model for Sr$_2$YRuO$_6$}

The above discussion of Sr$_2$YRuO$_6$ magnetic properties was based on the
molecular (cluster) picture, and we observed that oxygen plays a crucial
role in formation of the magnetic state. The same conclusion can be
obtained, starting from the extended band picture. The standard approach to
magnetism in the band theory goes back to Stoner and Slater\cite{ss}. They
considered non-interacting electrons in the paramagnetic state, and added
their exchange interaction in an average form, $H_{mag}=In_{\uparrow
}n_{\downarrow }=const-Im^2/4,$ where $m$ is total magnetization and $I$ is
independent of $m$. The magnetic susceptibility of such a system can be
written as 
\begin{equation}
\chi ^{-1}\equiv \partial ^2E/\partial n_{\uparrow }\partial n_{\downarrow
}=\chi _0^{-1}-I,  \label{stoner}
\end{equation}
where $\chi _0$ is Pauli susceptibility. If magnetization is measured in
Bohr magnetons, then $\chi _0=N(0),$ the density of states per spin at the Fermi
level. The instability occurs when $\chi $ diverges, that is,
when $IN(0)$ becomes larger than 1. Eq. \ref{stoner} can of course be viewed
as an approximation in the framework of the general linear response theory.
However, such an approximation is highly uncontrollable, and even the
splitting of the right-hand part of Eq. \ref{stoner} into two terms cannot
be derived in a systematic way. More instructive is application of the
Stoner method to the density functional theory. In DFT, total energy change
is exactly written as sum of the change in the one-electron energy, which
for small $m$ is $N(0)^{-1}m^2/4,$ and the change in the interaction energy,
which is ($\partial h/\partial m)m^2/4$ . Here $h=\langle V_{\uparrow
}-V_{\downarrow }\rangle $ is the effective Kohn-Sham magnetic field
averaged over the sample (because Stoner theory assumes a uniform internal
ferromagnetic field), and, $I\equiv -(\partial h/\partial m).$

The utility of the Stoner approach in DFT is due to the fact that usually
there are very few orbitals whose occupancy substantially influences $h,$
and therefore $I$ is easy to calculate in a quasiatomic manner, using, for
instance, the quasiatomic loop in standard LMTO codes. In practice, in
quasiatomic calculations one changes the occupation of a given orbital,
transferring some charge from the spin-up to spin-down quasiatomic level,
recalculates the LSDA potential and determines how large the induced
splitting of quasiatomic levels is.

If different kinds of atoms in a solid contribute to the density of states
at the Fermi level, one has to take into account the magnetization energy
for each of them. This means that the total Stoner $I$ for such a solid is
the average of the individual (quasiatomic) $I$'s with the squared partial
density of states. Indeed, suppose the states at the Fermi level are a
superposition of orbitals from several atoms, so that $N(0)=\sum_iN_i=N(0)%
\sum_i\nu _i$ (where $i$ labels the atoms). Applying a uniform magnetic
field creates a magnetization $m=\sum_im_i,$ where $m_i\equiv \nu _im$ is
magnetization of the $i$-th atom. By definition, the intraatomic energy
change is $-\sum_iI_im_i^2/4=-\sum_iI_i\nu _i^2m^2/4.$ Thus, the total $%
I=\sum_iI_i\nu _i^2.$

So formulated, the Stoner theory applies to infinitesimally small changes in
magnetization and essentially determines whether or not the paramagnetic
state is stable against ferromagnetism. It is, however, a reasonable
assumption that this theory holds, approximately, for finite magnetizations
as well. One has, however, to modify the one-electron energy term $%
N(0)^{-1}m^2/4,$ to account for the energy dependence of the density of
states, within the rigid-band approximation. Then, the spin splitting
producing a given magnetization $m$ can be defined as $\Delta =m/\bar{N}(m),$
where $\bar{N}(m)$ is the density of states averaged between the Fermi level
of the spin-up and spin-down subbands. For the one-electron energy one
obtains $\partial E_1/\partial m=m/2\bar{N},$ because one has to move $m/2$
electrons up by $\Delta .$ Integrating this expression, one arrives at the
so-called extended Stoner theory\cite{estoner}, which uses the following
expression for the total magnetization energy: 
\begin{equation}
E(m)=\frac 12\int_0^m\frac{m^{\prime }dm^{\prime }}{\bar{N}(m^{\prime })}-%
\frac{Im^2}4.  \label{StonerE}
\end{equation}
Minimization of this energy leads to the extended Stoner criterion, which
states that stable (or metastable) values of the magnetic moment are those
for which $\bar{N}(m)I=1$ and $d\bar{N}(m)/dm<0.$ The paramagnetic state is
(meta)stable when $\bar{N}(0)\equiv N(0)<1/I.$

Stoner theory is, in principle, formulated for a ferromagnetic instability.
However, unless the Fermi surface topology specifically favors (or
disfavors) the antiferromagnetic instability with a given vector {\bf Q},
one can assume that $\chi _0({\bf Q})\approx \chi _0(0).$ Indeed, in many
cases if a material comes out magnetic from the calculations,
the energy difference between ferro- and antiferromagnetic ordering is small
compared with the magnetic stabilization energy. As we shall see, this is
the case in Sr$_2$YRuO$_6,$ but not in SrRuO$_3,$ and the reason is that in
the latter the Stoner factor $I$ is very different for ferro- and
antiferromagnetic arrangements.

\begin{figure}[tbp]
\centerline{\epsfig{file=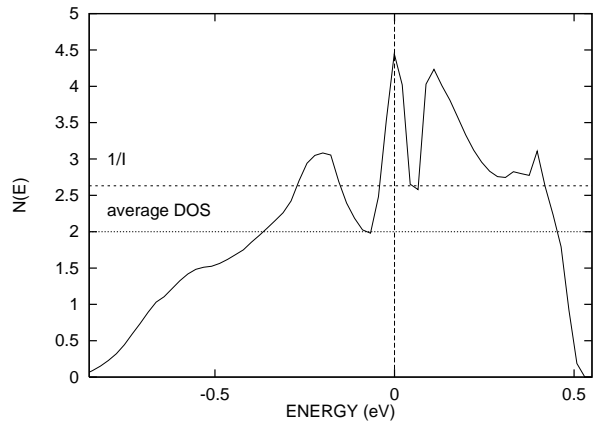,width=0.95\linewidth}}
\vspace{0.1in} \setlength{\columnwidth}{3.2in} \nopagebreak
\caption{LMTO density of states of the $T_{2g}$ band of the
nonmagnetic Sr$_2$YRuO$_6$, and inverse Stoner parameter $1/I$.}
\label{SYR-DOS-NM} \end{figure}

Now let us consider how one can describe the magnetism in Sr$_2$YRuO$_6$
from the Stoner point of view.
Calculation of Stoner parameters $I$'s is straightforward in the
LMTO method\cite{LMTO}, which divides space into atomic spheres. In the
popular Stuttgart LMTO-TB package it is possible to change occupancy of any
atomic orbitals and to calculate the resulting change in atomic parameters,
in particular the shift of the corresponding band center $C_{li}.$ With the
spin-up and spin-down occupancies split by $\pm m/2,$ the Stoner parameter
is ($C_{\uparrow }-C_{\downarrow })/m.$ We obtain $I_{Ru}$ of about 0.7 eV,
and, importantly, find that the O$_p$ states in ruthenates also have
substantial Stoner parameter, $I_O\approx 1.6$ eV. The density of the Ru
$d$ states is  approximately
twice larger than that of the three O $p$ states. Thus,  the total Stoner
parameter for Sr$_2$YRuO$_6$ is $I=I_{Ru}\nu _{Ru}^2+3I_O\nu _O^2\approx
0.38$ eV. Correspondingly, the paramagnetic state is unstable unless $N(0)
<2.6$ states/spin/eV/formula. The paramagnetic LMTO density of states of cubic
Sr$_2$YRuO$_6$ (that is, with breathing, but with no tilting distortion)
near the Fermi level is shown in Fig. \ref{SYR-DOS-NM}.
\begin{figure}[tbp]
\centerline{\epsfig{file=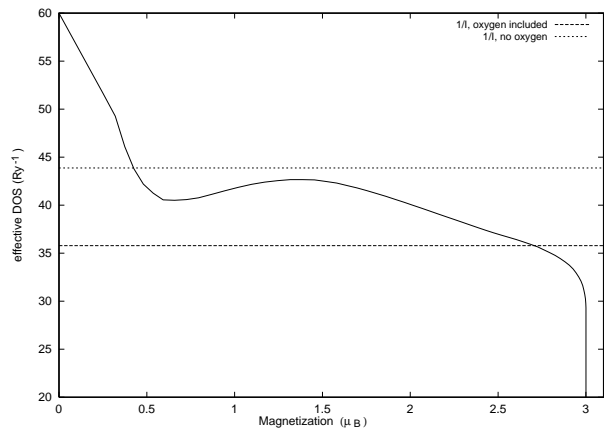,width=0.95\linewidth}}
\vspace{0.1in} \setlength{\columnwidth}{3.2in} \nopagebreak
\caption{Extended Stoner plot for the density of states shown
on Fig. \protect\ref{SYR-DOS-NM}} \label{SRYC-ST} \end{figure}

It has the narrow $T_{2g}$ band half filled, and $N(0)$ is close to 4.5
states/spin/eV/formula. This is much larger than $1/I$, so the paramagnetic
states is very unstable. On the other hand, the average 
density of states in the $T_{2g}$ band is not that
 large, $\tilde{N}\sim 3/W\approx 3/1.5
$ eV=2 states/spin$\cdot $eV, thus in the cubic structure this band will
not be fully polarized.

Integrating the density of states shown in Fig. \ref{SYR-DOS-NM}, 
we obtain the extended Stoner plot for Sr$_2$YRuO$_6$ (Fig.
\ref{SRYC-ST}), and observe that the equilibrium magnetization is slightly
smaller than 2 $\mu _B$, and the ground state is semimetallic, in agreement
with the self-consistent 
spin-polarized LMTO calculations, as well as with the more accurate
LAPW calculations.
Another fact that one can observe from Fig. \ref
{SRYC-ST} is that if oxygen would not contribute in the total Stoner factor,
that is, if total $I$ would be only $I_{Ru}\nu _{Ru}^2\approx
0.31$ eV, equilibrium magnetization would be very small, approximately 0.4 
$\mu _B$, and of course with a much smaller gain in energy. As we shall see
below, this is the case in SrRuO$_3$, where in the antiferromagnetic
structure oxygen ions cannot polarize by symmetry.

In Sr$_2$YRuO$_6$, however, oxygen fully contributes into the magnetic
stabilization energy both in ferro- and in antiferromagnetic structure, so
the next order mechanisms decides which magnetic order realizes.
Such an
additional mechanism is 
discussed in the previous section hybridization repulsion between the filled
and the empty $T_{2g}$ bands, which stabilizes
the antiferromagnetic structure.

\subsection{SrRuO$_3$ and CaRuO$_3$}

\subsubsection{Tight binding bands and relation to Sr$_2$YRuO$_6$.}
The main structural difference from the double perovskite, Sr$_2$YRuO$_6$ is
that now all oxygen ions are shared between two rutheniums, so one cannot
make use of a single RuO$_6$ cluster concept. As with Sr$_2$YRuO$_6,$ we
shall start by analyzing the band structure with non-tilted octahedra, that is,
with the cubic perovskite structure. Per cubic cell we have, in each spin
channel, two Ru $e_g$ states, strongly hybridized with 3 O $p_\sigma $
orbitals, and three Ru $t_{2g}$ states, hybridized with 6 O $p_\pi $
orbitals. In the nearest neighbor approximation, these $pd\sigma $ bands do
not mix with the $pd\pi $ bands, and the $pd\pi $ bands, in turn, consist of
three sets of mutually non-interacting $xy,$ $yz$, and $zx$-like bands. The
nearest neighbor TB Hamiltonians have the form
\end{multicols}
\rule[10pt]{0.45\columnwidth}{.1pt}
\[
H(e_g)=\left( 
\begin{array}{ccccc}
\ E_0(e_g) & 0 & 2t_\sigma s_x/\sqrt{3} & 2t_\sigma s_y/\sqrt{3} & 4t_\sigma
s_z/\sqrt{3} \\ 
0 & E_0(e_g) & 2t_\sigma s_x & -2t_\sigma s_y & 0 \\ 
2t_\sigma s_x/\sqrt{3} & 2t_\sigma s_x & E_0(p_\sigma )\  & 0 & 0 \\ 
2t_\sigma s_y/\sqrt{3} & -2t_\sigma s_y & 0 & E_0(p_\sigma )\  & 0 \\ 
4t_\sigma s_z/\sqrt{3} & 0 & 0 & 0 & E_0(p_\sigma )\ 
\end{array}
\right) 
\]
and

\[
H(xy)=\left( 
\begin{array}{ccc}
E_0(t_{2g}) & 2t_\pi s_x & 2t_\pi s_y \\ 
2t_\pi s_x & E_0(p_\pi ) & -4t_\pi ^{\prime }s_xs_y \\ 
2t_\pi s_y & -4t_\pi ^{\prime }s_xs_y & E_0(p_\pi )
\end{array}
\right) , 
\]
\begin{flushright}\rule{0.45\columnwidth}{.1pt} \end{flushright}
\begin{multicols}{2}
where $s_x=\sin (k_xa/2)$ etc. For each $t_{2g}$ manifold three bands
appear: one non-bonding at $E_0(p_\pi ),$ and one bonding-antibonding pair
at $E_{\pm }(xy)=0.5\{E_0(p_\pi )+E_0(t_{2g})\pm \sqrt{[E_0(p_\pi
)-E_0(t_{2g})]^2+16t_\pi ^2(s_x^2+s_y^2)}.$Analysis of the calculated band
structure shows that $E_0(t_{2g})\approx E_0(p_\pi ),$ so, neglecting
oxygen-oxygen hopping $t^{\prime },$ the dispersion is approximately $%
E_0(t_{2g})\pm 2t_\pi \sqrt{s_x^2+s_y^2},$ where $t_\pi \approx 1.4$ eV. Ru $%
e_g$ orbitals are split off from the $t_{2g}$ orbitals by about 3 eV. As in
Sr$_2$YRuO$_6,$ the crystal field effect on oxygen states is weaker: The O $%
p_\sigma $ states are less than 2 eV below O p$_\pi $ states. The energy
distance between Ru $e_g$ and O $p_\sigma $ levels is nearly 5 eV, so a good
approximation is $\Delta E=E_0(e_g)-E_0(p_\sigma )\gg t_\sigma .$ Applying
L\"{o}wdin perturbation theory to fold down the oxygen states, we get for
(antibonding) $E_g$ bands the effective Hamiltonian 
\end{multicols}
\rule[10pt]{0.45\columnwidth}{.1pt}
\[
H(e_g)=\left( 
\begin{array}{cc}
\ E_0(e_g)+4t_\sigma ^2(s_x^2+s_y^2+4s_z^2)/3\Delta E & 4t_\sigma
^2(s_x^2-s_y^2)/\sqrt{3}\Delta E \\ 
4t_\sigma ^2(s_x^2-s_y^2)/\sqrt{3}\Delta E & E_0(e_g)+4t_\sigma
^2(s_x^2+s_y^2)/3\Delta E
\end{array}
\right) , 
\]
\begin{flushright}\rule{0.45\columnwidth}{.1pt} \end{flushright}
\begin{multicols}{2}
which yields two bands with dispersion $\epsilon
_k=E_0(e_g)+8t^2(s_x^2+s_y^2+s_z^2\pm \sqrt{%
s_x^4+s_y^4+s_z^4-s_x^2s_y^2-s_z^2s_x^2-s_y^2s_z^2})/\Delta E.$

The formal valency of Ru in Sr$_x$Ca$_{1-x}$RuO$_3$ is 4. The total number
of electrons, populating the Ru-O valence bands, is 22. This means that the
bonding (mostly oxygen) $E_g$ bands are filled, as well as the bonding and
nonbonding $T_{2g}$ bands. The conduction band is the antibonding $T_{2g}$
band, with its 6 states filled by 4 electrons. This band has a strong
(logarithmic) van Hove singularity at half filling. However, direct
oxygen-oxygen hopping $t^{\prime }\approx 0.3$ eV, which we have initially
neglected, moves this singularity upwards to the position which corresponds
to approximately 63\% filling (3.8 electrons) and makes the singularity
sharper. This is the pronounced peak at $E_F$ in our first principles
paramagnetic DOS\cite{6}.
Such a situation, where the Fermi level nearly exactly hits a
logarithmic peak in the density of states, is energetically unfavorable, and
 leads to an
instability, which can be either magnetic, or
 a sufficiently strong lattice distortion, or both.

\subsubsection{Cubic perovskite: magnetic instability\label{SecSR3-mag}}

The calculated partial densities of Ru (d) and of the three 
O (p) states at the Fermi
level in Sr$_x$Ca$_{1-x}$RuO$_3$ are approximately
 70\% and 30 \%, respectively.
Correspondingly, $I=I_{Ru}\nu _{Ru}^2+3I_O\nu _O^2\approx 0.41$ eV. Without
the oxygen Stoner parameter, $I\approx 0.35.$ As mentioned above,
our LAPW calculations yield for SrRuO$_3$ in 
the cubic structure relatively small
magnetization of 1.17 $\mu_B$. The reason for that is that the density
of state is piled near the Fermi level, and drops quickly when one goes away
from it.  Fig. \ref{SR3+CR3-st}
shows how this is reflected in the effective density of states
$\tilde{N}(m)$: it decreases rapidly with magnetization, and becomes
equal to $1/I$ at $m\approx 1.2 \mu_B$. For a moderate tilting,
corresponding to actual  SrRuO$_3$ structure, $\tilde{N}(0)$ is
smaller than in the cubic structure, 
but it decreases rather slowly with $m$ and remains larger than  $1/I$ much
longer.

Two questions arise in this connection: why the ground state is
ferromagnetic, and not antiferromagnetic, and why in the actual
crystal structure is CaRuO$_3$ not magnetic
at all? The first question is particularly easy to answer. 
In an antiferromagnetic structure,  oxygen
ions occur between opposite spin Ru ions, and thus by symmetry have zero net
polarization. Correspondingly, the total Stoner parameter $I$ is smaller
and so is magnetic stabilization energy
and the equilibrium magnetization on Ru. As we shall see below,
tilting has a substantial effect on the effective density of states,
and for large tiltings  the ground state becomes paramagnetic. 
It follows from the above discussion, however, that 
the ground state is always either ferro- or paramagnetic.

\begin{figure}[tbp]
\centerline{\epsfig{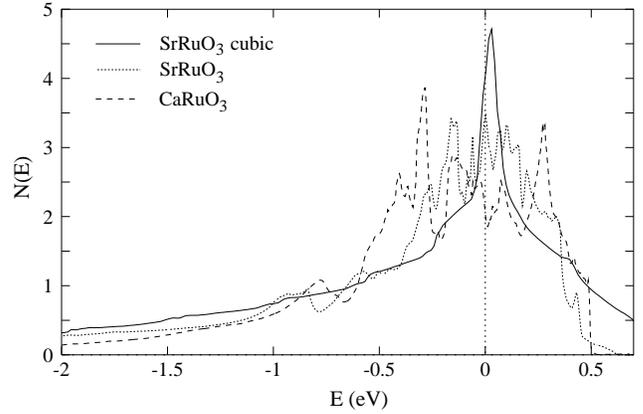}}
\vspace{0.1in} \setlength{\columnwidth}{3.2in} \nopagebreak
\caption{LAPW
 densities of states in the $T_{2g}$ band in  SrRuO$_3$ in the cubic
and its actual structure, and of CaRuO$_3$ in its actual structure,
and with experimental lattice parameters (4\% smaller for CaRuO$_3$).}
\label{SR3+CR3} \end{figure}
\begin{figure}[tbp]
\centerline{\epsfig{file=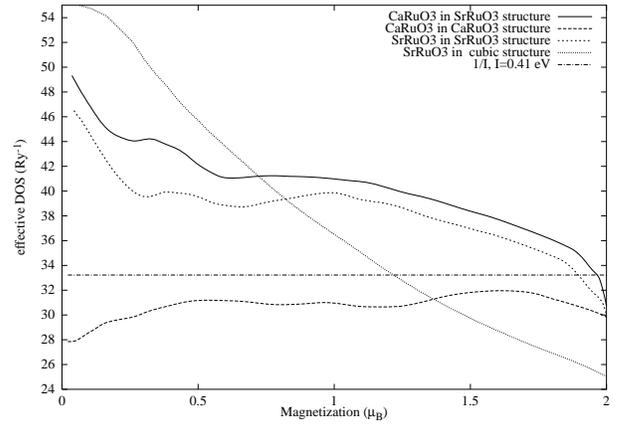,width=0.95\linewidth}}
\vspace{0.1in} \setlength{\columnwidth}{3.2in} \nopagebreak
\caption{Extended Stoner plot for SrRuO$_3$ and CaRuO$_3$ in various
structures, produced with the densities of states shown in Fig.
\protect\ref{SR3+CR3}. Inverse Stoner factors are calculated in the
LMTO atomic spheres as described in the text.}
\label{SR3+CR3-st} \end{figure}

This situation is in sharp contrast with classical localized magnetic
materials, like NiO or FeO, where even when ferromagnetism is imposed,
oxygen is polarized only very weakly, and the magnetization of the metal ion is
even smaller than for the antiferromagnetic state. It is also in contrast
with the ferromagnetic colossal magnetoresistance Mn oxides, where the
antiferromagnetic ground state is destroyed by the double exchange
interaction, competing with superexchange. These ruthenates have integer
occupancy of the valence band, and thus double exchange is not operative.
However, covalency effects, which are strong because of the large $pd$
hopping and the near-degeneracy of the Ru $(t_{2g})$ and O $(p_\pi )$
states, are operative. This strong covalency is what requires part of the
magnetic moment to reside on the oxygen, since exchange splitting the Ru $d$
states without the O would require disrupting the covalent bonding.

In the
crystal structures where it is possible to maintain O moments without
ferromagnetic ordering, an antiferromagnetic state is likely to form (as in
Sr$_2$YRuO$_6),$ but where it is not possible, like SrRuO$_3,$ a
ferromagnetic ground state occurs instead.
It is also worth noting that besides the double perovskite  Sr$_2$YRuO$_6,$
where oxygen ions can polarize both in the ferro- and antiferromagnetic
structure, and single perovskites  Sr$_x$Ca$_{1-x}$RuO$_3$, there
exist intermediate layered structures, which consist of perovskite
(Sr,Ca)O$_2$ layers. Using the same arguments, we
conjecture that if such compounds are
magnetic, the effect of oxygen will cause ferromagnetic ordering
inside layers, while interlayer coupling is strong ferromagnetic if the
layers are sharing apical oxygens, but may be antiferromagnetic if they are
connected by intermediate rocksalt layers (like in Sr$_2$RuO$_4$). 

\subsubsection{Role of the orthorhombic distortion}

The observed crystal structure of both SrRuO$_3$ and CaRuO$_3$ is
characterized by a substantial tilting of the RuO$_6$ octahedra. In SrRuO$_3$
the octahedra are rotated by 8$^{\circ },$ and in CaRuO$_3$ the distortion
is about twice larger. In Fig.\ref{SR3+CR3} we show the density of
states in the $T_{2g}$ band for these three different structures.
There are two interesting effects on the electronic
structure, associated with tilting. One is that hybridization between the $%
T_{2g}$ and $E_g$ bands becomes possible. This broadens the logarithmic
singularity in the density of states. At the same time the bands become more
narrow and the gap between the antibonding $T_{2g}$ and $E_g$ bands grows.
On the other hand, the unit cell is quadrupled so new Bragg reflections
appear. These yield pseudogaps at the new Brillouin zone boundaries,
occurring at energies close to half filling (e.g., along ${\bf \Gamma X}$ and 
${\bf \Gamma M}$ directions) as well at two-third filling (e.g., along $%
{\bf \Gamma R}$ direction). This second pseudogap thus appears to be near
the Fermi level. One factor, band narrowing, tends to increase the
equilibrium magnetization, but another one, the second pseudogap at the
Fermi level, works against it. The actual trend looks like this: at small
distortions the equilibrium magnetization grows. At some critical distortion
magnitude, which is not far from the observed equilibrium
distortion for SrRuO$_3$, the magnetization reaches a maximum and starts to
decline. The first principles 
calculations show little difference between SrRuO$_3$ and CaRuO$%
_3,$ provided the same crystal structure is used,
so the main difference in the observed behavior is indeed due to the
different distortion magnitudes.

\begin{figure}[tbp]
\centerline{\epsfig{file=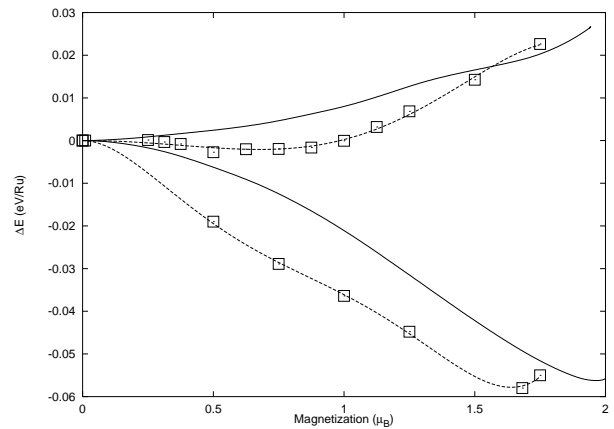,width=0.95\linewidth}}
\vspace{0.1in} \setlength{\columnwidth}{3.2in} \nopagebreak
\caption{Ferromagnetic stabilization energy  for CaRuO$_3$ 
in its actual crystal structure and in the SrRuO$_3$ crystal structure.
First-principle LAPW fixed-moment calculations (squares; the dashed
lines are guides to eye) are shown together
with the approximate Stoner formula (Eq.\protect\ref{StonerE}), based on
the data shown in  Fig. \protect\ref{SR3+CR3-st}.}
\label{fixedM} \end{figure}

To understand the changes caused by the tilting distortion it is instructive
to look at the extended Stoner plots for different distortions. Fig. \ref
{SR3+CR3-st} shows such plots for SrRuO$_3$ in the experimental structure, in
the cubic (ideal perovskite) structure, and for CaRuO$_3$, as well as for
CaRuO$_3$ in the SrRuO$_3$  structure. One may
immediately note the extreme instability of the cubic structure, due to
discussed peak at the Fermi level. However, because the density of states is
piled up near the Fermi level, the resulting exchange splitting is small
compared with the band width. For the moderate tilting, like that in the
experimentally observed SrRuO$_3$ structure, the peak broadens and it takes
larger exchange fields to fully split this peak into occupied and unoccupied
peaks. Finally, at even larger tiltings, corresponding to CaRuO$_3$, the
peak is suppressed. In the effective density of states, as shown on Fig.\ref
{SR3+CR3-st}, this results in a nearly flat plateau, extending from $m=0$ to $%
m\approx 1$ $\mu _B.$ Accidentally, this plateau matches nearly exactly $1/I,
$ calculated as described in the previous Section. In other words, the total
energy of CaRuO$_3$ is nearly independent of magnetization up to $m\approx 1$
$\mu _B!.$ The total energy as a function of magnetization is shown on Fig.%
\ref{fixedM}, where the results of the fixed-spin-moment LAPW calculations are
compared with the same energy differences in the Stoner theory\cite{notef}.

We conclude that although CaRuO$_3$ is nonmagnetic in its ground state,
long-wave paramagnons should be extremely soft in this compound. This
should effect the transport, magnetic and electronic properties.

\subsubsection{Transport properties}

Unusual transport properties of SrRuO$_3$ are due to the following three
peculiarities: (1) While the DOS in both spin subsystems are nearly the
same, the partial plasma frequency in the majority spin channel is 3 times
larger than in the minority spin channel, a manifestation of the proximity
to the half-metallic regime, which would occur if the magnetization were 2 $%
\mu _B$ instead of 1.59 $\mu _B.$ (2) There is strong coupling between
electrons, phonons, and magnons, which probably produces substantial
spin-flip scattering of electrons, and (3) In both spin channels the Fermi
surfaces consist of several sheets of complicated topology: hole-like,
electron-like, and open, so that hole-like and electron-like parts
compensate each other.

Let us start with the electric resistivity, and assume for simplicity that
two bands are present, spin-up and spin-down. Let us further assume that the
sources of the resistivity are scattering of electrons by phonons, with the
coupling constant $\lambda _{ph\uparrow \uparrow }=\lambda _{ph\downarrow
\downarrow }=\lambda _{ph},$ and by magnons, with the coupling constant $%
\lambda _m.$ Since the DOS are approximately equal, $N_{\uparrow
}=N_{\downarrow }=N\approx 23$ st/Ry,  also $\lambda _{\uparrow \downarrow
}=\lambda _{\downarrow \uparrow }=\lambda _m.$ The specific heat
renormalization in each band is now $(1+\lambda _{ph}+\lambda _m)\ $which
would need to be $\approx 4.0$ to agree with  experiment\cite{27}. In the
lowest-order variational solution of the Boltzmann equation, given by Allen%
\cite{pinski} (see also Ref. 
\onlinecite{eilat}), the resistivity of such a system at
sufficiently high temperature is $\rho =8\pi ^2kT(\lambda _{ph}+\lambda
_m)/\omega _p^2,$ where the so-called ``scattering-in'' term which is usually
small in cubic crystals is neglected, and $\omega _p^2=\omega _{p\uparrow
}^2+\omega _{p\downarrow }^2$ is the partial plasma frequency squared (one
can find in the literature \cite{fert} a so-called ``two-current formula''
which gives the same result when the ``scattering-in'' term is neglected.
There are some differences between the formulas of
Refs.\onlinecite{pinski} and  
\onlinecite{fert}, which we discuss in the Appendix.).

From our first principles calculations, $\omega _{p\uparrow }=3.3$ eV and $%
\omega _{p\downarrow }=1.5$ eV. In the nonmagnetic phase $\omega _p=6.2$
eV, the same as the
 total plasma frequency in the ferromagnetic phase. At $T\agt%
30$ K and up to the Curie temperature the resistivity is reported to be
linear\cite{27}. The linear coefficient ($\sim 1$ $\mu \Omega \cdot $cm/K)
corresponds to $(\lambda _{ph}+\lambda _m)\approx 2.9,$ very close to the
number extracted from the electronic specific heat. Above $T_C$ the
resistivity changes slope, remaining linear up to at least several hundred
Kelvin. The slope is, however, smaller than below $T_C,$ and corresponds to $%
\lambda =\lambda _{ph}+\lambda _{pm}\approx 1.5,$ where $\lambda _{pm}$ is
the electron-{\it para}magnon coupling constant. This value differs from
that quoted in Ref.\onlinecite{27}, because of considerable differences in the
calculated band structure. Thus, we conclude that the high-temperature
resistivity of SrRuO$_3$ indicates rather strong electron-paramagnon, and
even stronger electron-magnon coupling, with the reservation that probably
in this system one cannot really separate electron-phonon and
electron-magnon scattering completely because the corresponding degrees of
freedom are coupled. The problem noted in Ref.\onlinecite{27}, namely that at
high temperatures the mean free path is comparable to the lattice parameter,
yet no saturation is seen in the resistivity, remains.

The resistivity of CaRuO$_3$ has  also been studied. In the studies
reported in the literature \cite{1,13,bouch}the high-temperature
resistivity shows the same slope as in SrRuO$_3,$ in full agreement with our
observation that the electronic structure of both compounds is very similar.
At low temperatures, however, the resistivity behaves very differently,
namely it increases nearly linearly at small $T$ with a large slope. The
slope decreases eventually and at the room temperature the behavior becomes
similar to that of SrRuO$_3.$ This low-temperature linearity indicates that
the excitations responsible for resistivity (apparently, paramagnons) soften
at $T\rightarrow 0,$ indicating a magnetic instability at $T=0$ or at
slightly negative $T$ (in Curie-Weiss sense). This is in agreement with our
result that CaRuO$_3$ is on the borderline of a ferromagnetic state. Again,
paramagnons are strongly coupled with phonons, and this leads to the large
coupling strength.

The low-temperature resistivity has also attracted attention.
Experimentally, the resistivity initially increases rather quickly. Allen et
al \cite{27} observed a low-temperature power law $\rho (T)-\rho (0)\propto
T^{1-2}.$ Klein et al \cite{15} found that below 10 K the resistivity can be
reasonably well fit with a quadratic law, but an even better fit (linear),
was found for up to 30 K for the dependence of resistivity on magnetization.
We interpret this observation as follows: A stronger than $T^5$ increase
indicates that the excitations, responsible for the low-temperature
scattering, have a sublinear dispersion. Conventional magnons, with $\omega
\propto k^2$ dispersion, produce $\rho \propto T^2,$ in good agreement with
the experiment. In fact, the experimental exponent is even below 2, which
is easily accountable for by the Fermi surface effects: part of the
temperature dependence comes from the term $({\bf v}_{{\bf k}}-{\bf v}_{{\bf %
k}^{\prime }})^2,$ if it is proportional to $({\bf k}-{\bf k}^{\prime })^2;$
this not the case in SrRuO$_3,$ where one of the two Fermi surfaces
(majority spin) is a small sheet with heavy electrons (similarly, magnetic
alloys where momentum conservation does not hold show $\rho \propto T^{3/2};$
see Ref.
\onlinecite{fert32}). A possible problem with this interpretation of the
low-temperature resistivity is that as was already noted\cite{15},
in elemental ferromagnets, the magnon-limited resistivity is almost three
orders of magnitude smaller than what would be needed to explain the
low-temperature resistivity of SrRuO$_3$ (where $\rho \rightarrow \rho
_0+aT^2,$ $a\approx 0.02$ $\mu \Omega \cdot $cm/K$^2).$ This 
can be resolved if we invoke the anomalously large magnon-phonon coupling,
which, as discussed above, originates from the crucial role played by oxygen
in magnetic properties of ruthenates. The strong electron-magnon coupling at
low temperature in SrRuO$_3$ is closely related to the large
electron-paramagnon coupling at high temperatures and in CaRuO$_3.$ One can
make a rough estimate of the characteristic frequency of magnons responsible
for the resistivity: the Schindler-Rice formula\cite{rice}, derived for the $%
s-d$ paramagnon scattering should be qualitatively applicable here, because
we also have light electrons which carry current and are scattered by
magnons into a heavy, transport-inert band. This formula reads 
\begin{eqnarray*}
\rho (T) &\approx &\alpha (T/T_m)^2[J_2(T_m/T)-((T/T_m)^3J_5(T_m/T)] \\
J_n(x) &=&\int_0^x\frac{4z^ndz}{\sinh ^2(x/2)},
\end{eqnarray*}
and has asymptotic behavior at $T\rightarrow 0$ as $\alpha (T/T_m)^2\pi ^2/3$
and at $T\gg T_m$ as $\approx 0.8\alpha (T/T_m),$ where $kT_m$ is
characteristic energy of the 
magnons. Using the experimental number ($\rho -\rho
_0)/T^2\approx 0.02$ $\mu \Omega \cdot $cm/K$^2$ and assuming that the
magnon-limited part of the high-temperature resistivity is $\sim 0.5$ $\mu
\Omega \cdot $cm/K, we arrive at $T_m\sim 70$ K, which is a low, but not
impossible number.

The Hall coefficient in SrRuO$_3$
and CaRuO$_3$\cite{12} has attracted considerable attention.
 In both compounds the Hall constant $R$ shows an unusual
temperature dependence, changing sign at $T\sim 50$ K. At this point,
however, the similarity ends. For each given temperature the Hall
resistivity $\rho _{xy}$ in CaRuO$_3$ is nearly 
perfectly proportional to the
field, as it should be for ordinary Hall processes. In SrRuO$_3,$ to the
contrary, $d\rho _{xy}/dH$ decreases substantially with temperature, and only
well above $T_c$ does $\rho _{xy}$ become
 a linear function of $H.$ This closely
resembles the so-called extraordinary Hall effect in ferromagnets.

The physics of the extraordinary Hall effect is as follows: below $T_c,$
the internal magnetic field is much larger than that applied in a typical Hall
experiment. However, the Hall currents induced in different magnetic domains
mutually cancel. The applied field acts by  aligning domains and lifting this
cancellation. This process defines the large slope of $d\rho _{xy}/dH$ in
low fields. At a field close to the saturation magnetization $4\pi M_s$ all
domains are aligned and further change of the Hall current is due to the
applied field itself (the 
ordinary Hall effect). It is tempting to associate the
nonlinear field dependence of the Hall resistivity in SrRuO$_3$ with 
this effect. However, this
hypothesis has been discounted
 by the authors of Ref. \onlinecite{12} for  the
following reason: In  standard extraordinary Hall effect theory the
intersection of the linear low-field and high-field asymptotes occurs at $%
4\pi M_s.$ In SrRuO$_3$ the position of the intersection is roughly the same
for all temperatures below $T_c,$ and falls between 3 and 4 T. Magnetometer
data show that $4\pi M_s$ is about 0.12 T at $T=5$ K, and, naturally, drops
to zero at $T=T_c.$ Furthermore, a closer look at the data reveals that the
slope of the Hall coefficient
changes in a smooth manner, unlike conventional ferromagnets, where it
changes rather sharply near $H=$ $4\pi M_s.$  

Studies of the bulk magnetization
in polycrystalline\cite{7} and single crystal\cite{32}
 samples of SrRuO$_3$ show that the magnetization is not saturated
even at applied fields of several T. This has been ascribed to the strong
magnetocrystalline anisotropy, as measured by Kanbayashi\cite{32},
and expected
for a 4d magnet. Although hysteresis measurements for the thin film
samples on which Hall measurements were taken apparently showed saturation near
1.5 T, we speculate that the domains may not yet be fully aligned at this 
field, yielding a continuing non-linear field dependence in the Hall 
resistivity.

The sign reversal of the Hall
conductivity has received even more
 attention. In the literature two explanations can be found: one\cite{27}
assumes different temperature dependences for the electron and hole
scattering rates, because of the different scattering mechanism (phonon vs.
magnon), which may then yield a strong temperature dependence of the Hall
resistivity, and a sign change. It has been argued\cite{12} that
this hypothesis should not work, since CaRuO$_3$ is nonmagnetic, but still
shows a sign-changing Hall effect. Instead the authors of Ref. \onlinecite{12}
suggested that the sign may change because the number of electrons and holes
in the energy window $\sim kT$ around the Fermi energy may change with $T.$
However, the sign reversal in CaRuO$_3$ and SrRuO$_3$ could
be due to different physical reasons. This possibility is also suggested
by the very different field dependence of the Hall resistivity in the two
cases. On the other hand, it follows from our calculations, and is also
indicated by various experiments, that CaRuO$_3$ is on the verge of a magnetic
instability, and the interplay between the phonons and paramagnons may play
much the same role as the interplay between the phonons and magnons in SrRuO$%
_3.$ Furthermore, besides the temperature dependence of the relaxation rates
and the temperature broadening of the Fermi level, there is yet another
effect which may cause the sign change in SrRuO$_3$. The exchange
splitting must be very temperature dependent in SrRuO$_3.$ Unlike common
ferromagnets
 like Fe, where the Curie temperature corresponds to disordering
of  local moments, here the magnetization {\it disappears at T}$_c,$
including the {\it local} magnetization. Thus, the spin splitting changes
with the temperature, essentially disappearing around $T_{c.}$ This is in
contrast with most ferromagnets where an effective local
 spin splitting exists well
above $T_c,$ without any macroscopic magnetization. Thus, the band structure
itself is strongly temperature dependent. This effect can be operative in
SrRuO$_3$ in addition to the two other possible mechanisms.

\begin{figure}[tbp]
\centerline{\epsfig{file=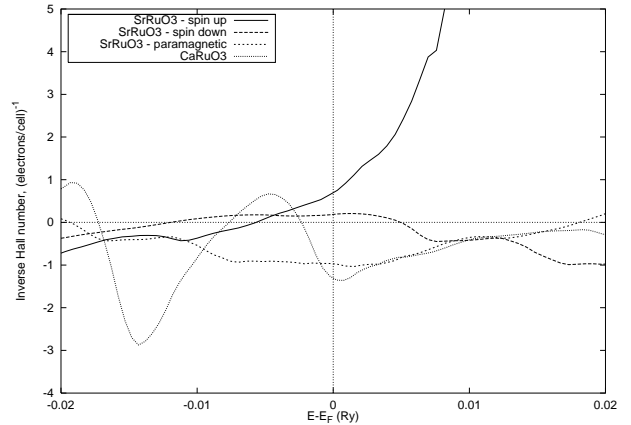,width=0.95\linewidth}}
\vspace{0.1in} \setlength{\columnwidth}{3.2in} \nopagebreak
\caption{Calculated inverse Hall number for SrRuO$_3$ (ferromagnetic
and nonmagnetic) and for  CaRuO$_3$. Note different signs for the two
spin subbands in  SrRuO$_3$ and strong dependence on the position
of the Fermi level.}
\label{hallSR3} \end{figure}

The prerequisite for any
of these mechanisms to be relevant is that there is strong compensation
between the hole-like and the electron-like contributions from the
different bands. To check this, we have calculated the Hall conductivity $%
\sigma _H$ and the Hall coefficient $R_H=\sigma _H/\sigma _0^2$ for all
the individual bands in SrRuO$_3$ and CaRuO$_3,$ following the procedure
described in Ref.\onlinecite{allenHall}. The results are shown in Fig. \ref
{hallSR3}. It was observed by Schultz {\it et al}\cite{allenHall} that
quantitative calculations of the Hall coefficient are extremely sensitive to
sampling of the Brillouin zone; it is impractical for these 20 atom
per unit cell structures to calculate the first principles band
structure at a {\bf k-}mesh comparable with the ultradense meshes used in Ref. 
\onlinecite{allenHall} for elemental metals and instead we have relied on 
interpolation between first principles band energies calculated at 100 points in the irreducible wedge of the zone. Thus, our calculations
shown in Fig. 
\ref{hallSR3} cannot be taken quantitatively, but rather illustrate the
qualitative fact, that the Hall conductivity has different signs in
different bands and spin channels. The net Hall conductivity is defined by
strong cancellation of hole- and electron-like contributions from different
bands sc which in turn is very sensitive to relative position of different
bands. Evidently, 
this balance can be easily violated by such temperature-dependent
factors as lattice distortion, magnetization, and relaxation times. The
mechanism suggested by Klein et al \cite{12} is also possible, since the net
Hall conductivity does change sign within a few
hundred K around $E_F$. Finally, very recent measurements\cite{guer}
 of the Hall
coefficient in mixed Sr$_x$Ca$_{1-x}$RuO$_3$ samples showed that for
intermediate concentrations it does not change sign with temperature,
suggesting that the sign reversals in pure compounds are accidental and
unrelated.

\section{Conclusions}

At this time there is already a fairly substantial body of experimental
literature on these ruthenates, including magnetic measurements,
spectroscopic studies, specific heat data and determinations of electronic
transport and superconducting properties. These measurements demonstrate
unusual and perhaps unexpected properties, and many of these have been
ascribed to correlation effects. For example, the specific heats in the
metallic compounds show substantial enhancements over the bare band
structure values, superconductivity occurs in a layered material in apparent
proximity to magnetic phases, quasiparticle bands measured by ARPES show
weaker dispersion than band structure calculations, satellites are
observed in angle integrated photoemission spectra, and the transport
properties of the metallic phases are unusual, showing e.g. sign reversals
in the Hall coefficient. Since this evidence clearly suggests something
unusual about the perovskite-derived ruthenates, it is tempting to ascribe
it  to strong correlation effects, particularly since these effects are all
either qualitatively in the direction expected for a correlated system or
can conceivably arise from the additional complexity introduced by
correlation effects.

On the other hand, chemical trends lead to the expectation that, all things
being equal, 4d Ru oxides should be less prone to strongly correlated
behavior than the corresponding 3d oxides, and much less prone to such
effects than cuprates. This is because of the much more extended 4d orbitals
in Ru ions which should lead to stronger hybridization, better screening and
lower effective Hubbard $U$. Furthermore, although  much of the data are at
first sight qualitatively in accord with general expectations for a
correlated system, they have not been quantitatively explained in these
terms, and there are some data that are rather difficult to understand 
purely in terms of a correlated scenario, most notably the disappearance of
magnetism upon doping Ca in the 
SrRuO$_3$ system, and the ferromagnetic ground state
in the integer occupancy (and thus not a double-exchange system) compound,
SrRuO$_3$.

We have performed first principles, band structure based calculations within
the LSDA for SrRuO$_3$, CaRuO$_3$ and Sr$_2$YRuO$_6.$ Although this approach
fails miserably in systems that are truly strongly correlated, it does yield
the correct magnetic and electronic states in these materials,
including quantitative agreement
with known magnetic properties in all cases in these ruthenates. Moreover, the
different magnetic behaviors can be fully understood in terms of  simple and
straightforward one-electron tight binding models and Stoner theory.
Although interpretation of the transport properties in terms of a conventional
one-electron picture and Bloch-Boltzmann theory is not as straightforward, we
show that such an approach is not inconsistent with the existing
body of experimental
evidence. A key notion for understanding the transport in these systems is
strong electron-phonon-(para)magnon coupling, which in turn can be
understood in the framework of the band theory. 

In agreement with expectations based on chemical trends, rather strong
hybridization is found between the Ru 4d and O 2p states in these materials.
While antagonistic to a strong correlation scenario this is in large part
responsible for the unusual properties in our band picture,
including the very fact that
magnetism occurs at all in a 4d metallic oxide. This strong hybridization
leads to a ferromagnetic direct exchange interaction between Ru and O, and
the cooperation between Ru and O contributions to the Stoner parameter leads
to the magnetic ground states. As a result, the O ions in these ruthenates
make substantial contributions to the magnetization density, which may be
observable in neutron scattering experiments with O form factors included in
the refinements. The importance of $p-d$ hybridization also leads to a
strong coupling of magnetic and structural degrees of freedom, resulting in
for example the destabilization of the ferromagnetic state due to octahedral
tilting in CaRuO$_3$.

One consequence of our scenario is that when Ru ions are bonded to the same
O, as neighboring Ru ions are in the perovskite structure, the interaction
between them will be strongly ferromagnetic. This means that magnetic
fluctuations in layered ruthenates like Sr$_2$RuO$_4$ and the associated
Ruddlesden-Popper (RP) series of compounds, are predicted to have
predominantly ferromagnetic character in plane, although alternating layers,
or perovskite blocks in the RP series may  be coupled
antiferromagnetically to each other
via superexchange through rocksalt blocks. Such
ferromagnetic fluctuations would be pair-breaking for singlet ($s-$ or $d-$%
wave) superconductivity, but not for triplet superconductivity, as
suggested for instance by  Rice and Sigrist\cite{26} for Sr$_2$RuO$_4$. In
fact, for triplet pairing both magnetic fluctuations and phonons provide
Cooper attraction. Finally, when Ru ions are not connected at all via common
O ions, like in Sr$_2$YRuO$_6$
 the Ru-Ru coupling is via two intervening O ions, both of
which are strongly hybridized with and ferromagnetically coupled to the
nearest Ru, but couple to each other via an antiferromagnetic superexchange
interaction. This results in an antiferromagnetic state.

The strength and importance of covalent transition metal - oxygen
interactions, combined with magnetism and metallicity is perhaps unique to
these ruthenates. Already a number of interesting physical properties have
been found among these compounds, and no doubt more interesting physics
remains to be found in this family.

\acknowledgements

We acknowledge enlightening discussions with J.S. Dodge, R.P. Guertin,
Lior Klein, Mark Lee, and W.E. Pickett. Work at the
Naval Research Laboratory is supported by the Office of Naval Research.
Computations were performed using the DoD HPCMO computing centers at NAVO
and ASC.

\appendix
\section{}

Introduced by Fert and Campbell \cite{fert} and often used since then, the 
two-current conduction model of transport in ferromagnetic metals assumes that
the spin-up and spin-down distribution function changes are independent.
This formula was derived for a very specific case of $s-d$ scattering in
transition metals and under a number of simplifying assumptions.
In the original papers \cite{fert} no
clear distinction was made between the spin-flip scattering (which always
influences the total conductivity) and the spin-flip conductivity (which
vanishes when scattering-in is neglected). In a later publication \cite
{fertbook} this distinction has been made, but this publication is less
well known. We find it instructive to present here a systematic derivation for
the two-current model of electric transport in ferromagnets, and to show how
it is related to the general theory of multiband conductivity. This
derivation is based on Allen's variational approach to multiband
Boltzmann equation\cite{allenFSH,pinski} and clearly shows some
 limits to the
applicability of the two-current model.

The Boltzmann equation for electric transport in metals is 
\begin{eqnarray*}
e{\bf E\cdot v}_{i{\bf k}}\frac{\partial f}{\partial \epsilon _{i{\bf k}}}%
&=&\sum_{j{\bf k}^{\prime }}P_{i{\bf k,}j{\bf k}^{\prime }}(\delta f_{i{\bf k}%
}-\delta f_{j{\bf k}^{\prime }})\\
&=&\delta f_{i{\bf k}}\sum_{j{\bf k}^{\prime
}}P_{i{\bf k,}j{\bf k}^{\prime }}-\sum_{j{\bf k}^{\prime }}P_{i{\bf k,}j{\bf %
k}^{\prime }}\delta f_{j{\bf k}^{\prime }} ,
\end{eqnarray*}
where the subscripts $i,j$ include both the band index and the spin and $P$
is the transition probability matrix element, which can also be written
conveniently in terms of the scattering matrix elements $M$ as 
\[
P_{i{\bf k,}j{\bf k}^{\prime }}=M_{i{\bf k,}j{\bf k}^{\prime }}^2\delta
(\epsilon _{i{\bf k}}-\epsilon _{j{\bf k}^{\prime }})
\]
for static (impurity) scattering, or the corresponding expression for the
scattering by phonons, magnons etc. The change of the distribution function
for given {\bf k} is $\delta f_{i{\bf k}}=-\phi _{i{\bf k}}(\partial
f/\partial \epsilon _{i{\bf k}}),$ so that 
\end{multicols}
\rule[10pt]{0.45\columnwidth}{.1pt}
\[
e{\bf E\cdot v}_{i{\bf k}}\frac{\partial f}{\partial \epsilon _{i{\bf k}}}
=\sum_{j{\bf k}^{\prime }}P_{i{\bf k,}j{\bf k}^{\prime }}(\frac{\partial f%
}{\partial \epsilon _{i{\bf k}}}\phi _{i{\bf k}}-\frac{\partial f}{\partial
\epsilon _{j{\bf k}^{\prime }}}\phi _{j{\bf k}^{\prime }}) 
\] and for $T\rightarrow 0$ \begin{eqnarray*}
e{\bf E\cdot v}_{i{\bf k}}\delta (\epsilon _{i{\bf k}}-E_F) &=&\sum_{j{\bf k}%
^{\prime }}M_{i{\bf k,}j{\bf k}^{\prime }}^2(\phi _{i{\bf k}}-\phi _{j{\bf k}%
^{\prime }})\delta (\epsilon _{i{\bf k}}-\epsilon _{j{\bf k}^{\prime
}})\delta (\epsilon _{i{\bf k}}-E_F)\\
&=&\sum_{j{\bf k}^{\prime }}M_{i{\bf k,}j%
{\bf k}^{\prime }}^2(\phi _{i{\bf k}}-\phi _{j{\bf k}^{\prime }})\delta
(\epsilon _{j{\bf k}^{\prime }}-E_F)\delta (\epsilon _{i{\bf k}}-E_F)
\end{eqnarray*}
Allen introduced (and called ``disjoint representation'') the following
approximation: for each sheet of the Fermi surface $\phi _{i{\bf k}}$ is
proportional to {\bf v}$_{i{\bf k}},$ $\phi _{i{\bf k}}=a_i{\bf v}_{i{\bf k}%
}\cdot e{\bf E},$ where the coefficients $a$ depend on the band (and, in a
magnet, on the spin). In this approximation, the last equation may be
solved (we assume {\bf E}$\parallel x$ and omit subscript $x$ at $v):$ 
\[
\sum_{i{\bf k}}v_{i{\bf k}}^2\delta (\epsilon _{i{\bf k}}-E_F)=\sum_{i{\bf k,%
}j{\bf k}^{\prime }}M_{i{\bf k,}j{\bf k}^{\prime }}^2(a_iv_{i{\bf k}}^2-a_j%
{\bf v}_{i{\bf k}}\cdot {\bf v}_{j{\bf k}^{\prime }})\delta (\epsilon _{j%
{\bf k}^{\prime }}-E_F)\delta (\epsilon _{i{\bf k}}-E_F)
\]
or 
\[
\langle Nv^2\rangle _i=\sum_{j}Q_{ij}a_j,
\]
where the shorthand notation on the left-hand side is obvious and 
\begin{eqnarray*}
Q_{ij} &=&\delta _{ij}\sum_j\sum_{{\bf kk}^{\prime }}M_{i{\bf k,}j{\bf k}%
^{\prime }}^2v_{i{\bf k}}^2\delta (\epsilon _{i{\bf k}}-E_F)\delta (\epsilon
_{j{\bf k}^{\prime }}-E_F)-\sum_{{\bf kk}^{\prime }}M_{i{\bf k,}j{\bf k}%
^{\prime }}^2{\bf v}_{i{\bf k}}\cdot {\bf v}_{j{\bf k}^{\prime }}\delta
(\epsilon _{i{\bf k}}-E_F)\delta (\epsilon _{j{\bf k}^{\prime }}-E_F) \\
&=&\delta _{ij}\sum_j\langle NM^2Nv^2\rangle _{ij}-\langle NvM^2Nv\rangle
_{ij}.
\end{eqnarray*}

\begin{multicols}{2}
The first term here accounts for the scattering-out, and the second for the
scattering-in processes. Solving now for $a$'s, we find 
\[
a_i=\sum_jQ_{ij}^{-1}\langle Nv^2\rangle _j.
\]
Since the total current density is 
\begin{eqnarray*}
{\bf j}&=&-2e\sum_{i{\bf k}}{\bf v}_{i{\bf k}}\phi _{i{\bf k}}\delta (\epsilon
_{i{\bf k}}-E_F)\\
&=&-2e^2\sum_{i{\bf k}}v_{i{\bf k}}^2\delta (\epsilon _{i{\bf k%
}}-E_F)a_i{\bf E=-}2e^2\sum_j\langle Nv^2\rangle _ia_i{\bf E},
\end{eqnarray*}
and the conductivity is
\[
\sigma =e^2\sum_{ij}\langle Nv^2\rangle _iQ_{ij}^{-1}\langle Nv^2\rangle _j.
\]
In notations of Fert and Campbell\cite{fert,fertbook} this is
\[
\sigma =\sum_{ss^{\prime }}\rho _{ss^{\prime }}^{-1},
\]
analogous to Eq. 57 of Ref.\onlinecite{pinski}. Campbell-Fert formula
is different when the nondiagonal elements are not neglected. It gives instead
\[\sigma =(\rho _1+\rho_2+4 \rho_{12})/(\rho _1 \rho_2+\rho _1  \rho_{12}+
\rho_2  \rho_{12}).
\]
 
For the purpose of this paper we shall neglect scattering-in completely,
because it is defined by an average over the sign-changing quantity ${\bf v}%
_{i{\bf k}}\cdot {\bf v}_{j{\bf k}^{\prime }}$, so that the matrix $Q$ is
diagonal. We shall also define the partial resistivities differently from
Ref. \onlinecite{fert,fertbook}, namely so that 
\[
\rho _{ij}=\langle Nv^2M^2\rangle _{ij}/2e^2\langle Nv^2\rangle _i^2=\frac 1{%
e^2\tau _{ij}}/\left( \frac nm\right) _i^{eff}.
\]
So defined, the $\rho _{\uparrow \downarrow }$ is proportional to the
spin-flip scattering rate. Then, for a simple ferromagnet with the two Fermi
surface sheets, one for spin-up and another for spin-down electrons, we have 
\[
\sigma =(\rho _{\uparrow }+\rho _{\uparrow \downarrow })^{-1}+(\rho
_{\downarrow }+\rho _{\downarrow \uparrow })^{-1}
\]
or 
\[
\rho =(\rho _{\uparrow }\rho _{\downarrow }+\rho _{\downarrow }\rho
_{\uparrow \downarrow }+\rho _{\uparrow }\rho _{\downarrow \uparrow }+\rho
_{\uparrow \downarrow }\rho _{\downarrow \uparrow })/(\rho _{\uparrow }+\rho
_{\downarrow }+\rho _{\uparrow \downarrow }+\rho _{\downarrow \uparrow }).
\]
This coincides with the Fert-Campbell formula, if the scattering-in term is
neglected; otherwise, it is different.

\end{multicols}

\begin{references}
\bibitem{notef}  Note that we have not optimized the structure either
 for the paramagnetic or for the magnetic structures in our fixed-moment
 calculations. Due to strong magnetoelastic interactions the equilibrium 
 tilting angle in a magnetic state should be slightly different from
 that without magnetization. However, the qualitative statement that CaRuO$_3$
 is on the verge of magnetic instability should be true even if
 magnetoelastic effects are taken into account.
\end{references}
\end{document}